\documentclass[amsmath,amssymb,12pt]{article}
\usepackage[all]{xy}
\usepackage[centertags]{amsmath}
\usepackage{latexsym}
\usepackage{amsfonts}
\usepackage{amssymb}
\usepackage{amsthm}
\usepackage{graphicx}
\usepackage{dcolumn}
\usepackage{bm}

\begin{document}




\title{Numerical Analysis of Three-dimensional Acoustic Cloaks and Carpets}


\author{Guillaume Dupont$^{1}$, Mohamed Farhat$^{2}$, Andr\'e Diatta$^{3},$\\ S\'ebastien Guenneau$^{1}$ and Stefan Enoch$^{1}$}


\maketitle
\date{}
\noindent
{\footnotesize
\thanks{\centerline{$^{1}$Institut Fresnel-CNRS (UMR 6133), Aix-Marseille Universit\'e}
\newline\noindent
 \centerline{Campus universitaire Saint-J\'er\^ome, 13013
Marseille, France.}\\
 \centerline{Email addresses: guillaume.dupont@fresnel.fr; }
\\
\centerline{sebastien.guenneau@fresnel.fr; stefan.enoch@fresnel.fr}
\\
\centerline{$^{2}$ Department of Electrical and Computer Engineering}\\
\centerline{The University of Texas at Austin, Austin, TX, 78712, USA}
\\
 \centerline{Email address: mohamed.farhat@fresnel.fr }
\\
\centerline{ $^{3}$University of Liverpool. Department of Mathematical Sciences,}
\\
\centerline{ M.O. Building, Peach Street, Liverpool L69 3BX, UK}
\\
 \centerline{Email address: adiatta@liv.ac.uk;}
 }


\begin{abstract}
We start by a review of the chronology of mathematical results on the Dirichlet-to-Neumann map
which paved the way
towards the physics of transformational acoustics. We then rederive the expression for
the (anisotropic) density and bulk modulus appearing in the pressure wave equation written
in the transformed coordinates.
A spherical acoustic cloak consisting of an alternation of homogeneous
isotropic concentric layers is further proposed based on the effective medium theory.
This cloak is characterised by a low reflection and good efficiency over a large bandwidth for both near and far fields, which
approximates the ideal cloak with a inhomogeneous and anisotropic distribution of
material parameters. The latter suffers from singular material parameters on its inner surface. This
singularity depends upon the sharpness of corners, if the cloak has an irregular boundary, e.g. a polyhedron
cloak becomes more and more singular when the number of vertices increases if it is star shaped.
We thus analyse the acoustic response of a non-singular spherical cloak designed by blowing up a small
ball instead of a point, as proposed in [Kohn, Shen, Vogelius, Weinstein, Inverse Problems 24, 015016,
2008]. The multilayered approximation of this cloak requires less extreme densities (especially
for the lowest bound). Finally, we investigate another type of non-singular
cloaks, known as invisibility carpets [Li and Pendry, Phys. Rev. Lett. 101, 203901, 2008], which mimic
the reflection by a flat ground.
\end{abstract}







\section{Introduction: Dirichlet-to-Neumann map versus cloaking}
\label{intro}

Several mathematical studies of invisibility devices have emerged in the early eighties from a consideration of electric impedance tomography (EIT) models \cite{kohn84a,kohn84b,uhlmann,lassas}.
The aim of this technique is to probe the spatial variations in conductivity $\sigma$
within some three-dimensional bounded region $\Omega$ by applying a known static voltage $u$ to the surface $\partial\Omega$
bounding the region and recording the resulting current $\sigma\nabla u$.
The equation satisfied by these quantities inside $\Omega$ is
\begin{equation}
\nabla\cdot(\sigma\nabla u)=0 \; .
\label{eqcondu}
\end{equation}
The current-voltage relationship provides a Dirichlet-to-Neumann map $\Lambda_\sigma$ on $\partial\Omega$:
\begin{equation}
\Lambda_\sigma : u\mid\partial\Omega\longmapsto (\sigma\nabla u)\cdot{\bf n}\mid\partial\Omega \; ,
\label{trace}
\end{equation}
where ${\bf n}$ is the unit outward normal to the surface $\partial\Omega$.
In order for EIT to work, it must
be possible to determine $\sigma$ from a knowledge of $\Lambda_\sigma$.
If this can be done, then cloaking (in this situation) is impossible, even in theory. The question of
whether or not the mapping can be used to determine the form of the conductivity is known
as the Calderon problem \cite{calderon}. It turns out \cite{kohn} that the
Dirichlet-to-Neumann map does determine $\sigma$ , but only under certain conditions: namely, that $\sigma$ must be known to be scalar-valued, positive and finite
\footnote{It is generally assumed in the physics literature that $\sigma$ is a smooth function, with regularity depending on the dimension of the problem. In dimension 2, however, mathematicians have established that no smoothness is required: only boundedness and positivity are needed \cite{Astala2006}.}.
However, if  some of these conditions are not met (for instance $\sigma$ is matrix valued), cloaking may be possible.
The singularity and
anisotropy of the material parameters required for such a conductivity cloak allow for the possibility that
the Dirichlet-to-Neumann map does not uniquely determine the conductivity \cite{syl}.
In 2006, Leonhardt proposed a design for a two-dimensional electromagnetic cloak
which works in the geometric optics limit. Here, the propagation of light is governed
by the Helmholtz equation, for which a similar uniqueness theorem has been established using tools
of quasi-conformal maps \cite{leonhardt06}.
Electromagnetic sensing in the quasistatic
limit, which is closely related to cloaking, was reported
by the mathematician Lassas back in 1997 \cite{lassas}. The analysis
of Lassas, based on variational principles (see also \cite{g1} and \cite{onofrei} for the all-frequency case)
uses the fact that mathematically, the trace
map in (\ref{trace}) associated with the Helmholtz equation
$\nabla\cdot(\sigma\nabla u)=\omega^2 u$
is well defined and invertible from the Hilbert space $H^{-1/2}(\Omega)$ to the Hilbert space $H^{1/2}(\partial\Omega)$
provided that the conductivity $\sigma$ is a uniformly elliptic symmetric-matrix-valued function and
the spectral parameter $\omega^2$ avoids a discrete set of eigenvalues of the Neumann Laplacian (which can be interpreted as eigenfrequencies
associated with almost-trapped eigenstates $u$ \cite{g2} in the context of quantum mechanics): This comes from the
fact that the aforementioned Helmholtz equation subject to the boundary condition $\partial u/\partial n=f$ forms
a well-posed boundary value problem and ${\Vert u \Vert}_{H^1(\Omega)}\leq C{\Vert f \Vert}_{H^{-1/2}(\partial\Omega)}$ where $C$ is a strictly positive constant depending upon $\Omega$ (assumed to be
smooth enough, which excludes the case of bad domains, such as star-shaped as numerically studied in section
6 of this paper).
Actually, a singular change of variable resulting in perfect cloaking in the context of
electric impedance tomography was introduced by Greenleaf et al
in space dimension $n\geq 3.$ This was a variation of an abstract  two-dimensional example
of Lassas, et al \cite{lassastaylor2003}.
 Kohn et al show in \cite{kohn} that
the situation is not significantly different when $n = 2$: perfect
cloaking is also possible in space dimension two, using tools
for the removability of singularities for harmonic functions.
In the resonance domain, whereby the wavelength of the wave incident on the cloak is about
the same size as its diameter, it is actually necessary to use vector Maxwell's equations to
account for effects of polarization and this was done back in 2006 by
Pendry, Schurig and Smith \cite{pendry}. Although this last paper did not contain
a mathematically clean derivation of the problem, a subsequent experimental paper involving the same authors
\cite{pendryexp} was convincingly demonstrating that cloaking is possible, at least for microwaves,
and at a specific frequency, 8.5 GHz. A full mathematical account of the invisibility problem in
electromagnetism can be found in a paper by Greenleaf, Kurylev, Lassas and Uhlmann \cite{g1}, which is
written with a strong mathematical flavor.

In \cite{kohn84a}, it is shown (with a credit therein to Luc Tartar) that the essence of indistinguishable media \footnote{We avoid the word {\it cloaking} here as
its meaning varies significantly between authors: for instance, Kohn et al. consider in \cite{kohn} that cloaking occurs when two different media have the same boundary measurements, while Greenleaf et al. further require in \cite{g1} that one medium contains a cavity in which
general objects can be hidden. The latter definition corresponds to invisibility cloaks designed by physicists, see e.g. \cite{pendry,pendryexp}, and seems to require a singular transformation $F$, although whether this is really mathematically necessary is still unknown for $n\geq 3$. Another commonly used terminology is {\it external cloaking} which refers to a countable set of dipoles hidden by the presence of a resonant object, see \cite{Ross_cloaking,milton3}.} is encompassed in boundary measurements
in a very simple way. We find it is worthwhile reproducing the main argument here:
Let $F$ denote the map from the original set of coordinates to the new one.
If $F(x) = x$ at $\partial\Omega$, then the boundary measurements
associated with $\sigma$ and $\tilde\sigma=DF(x)\sigma(x)DF(x)^T/det(DF(x))
:={\bf J}\sigma{\bf J}^T/det({\bf J})$ are identical,
keeping in mind the coefficients $J^i_j=\partial y_i/\partial x_j$ of the Jacobian matrix
are evaluated at point $x=F^{-1}(y)$. In other words, we are
assured that $\Lambda_\sigma(f)=\Lambda_{\tilde\sigma}(f)$ for all $f$ Dirichlet data of (\ref{eqcondu}).

The proof of this statement \cite{kohn} is straightforward since if $F(x) = x$ at $\partial\Omega$,
then the change of variables $F$ does not affect the Dirichlet data. If $\sigma(x)$ is bounded positive
and definite, then the solution of (\ref{eqcondu}) with Dirichlet data $f$ is given by the minimum of the
energy functional $\int_\Omega \sigma(x)\nabla_x u \cdot \nabla_x u \, dx$, so for
any $f$, we have
\begin{equation}
\int_{\partial\Omega} f\Lambda_\sigma f \, dx
= \min_{u=f \hbox{ at } \partial\Omega} \int_\Omega \sigma(x)\nabla_x u \cdot \nabla_x u \, dx
=  \min_{u=f \hbox{ at } \partial\Omega} \int_\Omega \tilde\sigma(y)\nabla_y u \cdot \nabla_y u \, dy
= \int_{\partial\Omega} f\Lambda_{\tilde\sigma} f \, dy \; .
\end{equation}
Importantly, this means the Dirichelt-to-Neumann map $\Lambda_\sigma$ can determine $\sigma$ at best
up to a change a variables, as first noted by Kohn and Vogelius in 1984
\cite{kohn84a} with a reference therein to Luc Tartar: $\Lambda_\sigma$ and $\Lambda_{\tilde\sigma}$
determine identical quadratic forms on Dirichlet data. This is the essence of cloaking.
If, however, $\sigma$ is scalar-valued, positive,
and finite, then the situation is very different: under some conditions
on the regularity of $\Omega$, knowledge
of the Dirichlet-to-Neumann map $\Lambda_\sigma$ determines an internal isotropic conductivity
$\sigma$ uniquely, and cloaking breaks down. The salient consequence is that any
cloaking device requires anisotropy, a fact known since the
eighties, due to Sylvester and Uhlmann \cite{syl} for $n\ge 3$ in 1987 and Nachman \cite{nachman95} for $n=2$ in the 1995, but rediscovered and popularized twenty years afterwards in the context of invisibility cloaks by Pendry and his colleagues.

In the present paper, we are actually interested in pressure waves propagating in
fluids. Such a problem is of a different physical nature than electromagnetic waves interacting with invisibility cloaks
(e.g. there are no polarization effects),
but the associated governing equation (the scalar Helmholtz equation) is a simplified version of Maxwell's equations \cite{g1}. This might explain why acoustic
cloaking has been touched upon by various photonics groups including \cite{cummernjp,sanchez} in two-dimensional settings and \cite{chen07,pendryprl} in three-dimensional settings.
Using tensors notations with Einstein's summation convention, one can write down the form-preserving transformations for the density and bulk modulus as:
\begin{equation}
\underline{\underline{\tilde{\sigma}}}^{ij}=\frac{J^i_k \underline{\underline{{\sigma}}}^{kl} J^j_l} {det(J^i_k)}
\; , \;
\tilde{\kappa} = \dfrac{\kappa}{det(J^i_k)} \; ,
\label{vip}
\end{equation}
with $\underline{\underline{\tilde{\sigma}}}$ indicating that it is a rank-2 tensor. This very useful formula is rederived
using simple arguments in section 2.

We emphasize that the formula (\ref{vip}) associated with transformation based solutions to the conductivity and acoustic equations is a version of the form preserving transformation for the tensors of permittivity and permeability
$\underline{\underline{\tilde{\epsilon}}}^{ij}=(J^i_k \underline{\underline{{\epsilon}}}^{kl} J^j_l)/det(J^i_k)$
and $\underline{\underline{\tilde{\mu}}}^{ij}=(J^i_k \underline{\underline{{\mu}}}^{kl} J^j_l)/det(J^i_k)$
appearing in  Maxwell's equations in curvilinear coordinate systems,
which were used by Pendry, Schurig, and Smith \cite{pendry}
in order to detour electromagnetic waves around a spherical region of space, with subsequent
work on arbitrarily sized and shaped solids by Nicolet et al. \cite{opl2008}.
As already mentioned, the experimental validation of these theoretical
considerations was given by Schurig et al. \cite{pendryexp}, who used
a cylindrical cloak consisting of concentric arrays of split
ring resonators which made a copper cylinder invisible to
an incident plane wave at 8.5 GHz (via artificial magnetic activity).
However, in order to
broaden the range of frequencies over which a cloaking
metamaterial works, one needs to explore other routes
avoiding resonant elements. Huang et al. proposed
a concentric multilayered design of cylindrical electromagnetic cloaks \cite{huang}.
Farhat et al. further divided these layers in infinitely
conducting sectors in the context of electromagnetics \cite{farhatoe}.
A very comphensive, yet rigorous, mathematical account of the homogenization approach to
cloaking can be found in a physics paper by Greenleaf et al. \cite{g2}, while numerical
illustrations of this theory are discussed in Farhat et al. \cite{farhatnjp}.

There are nevertheless genuinely non-singular routes towards cloaking:
in 2008 Pendry and Li proposed to design an invisibility carpet so that
a bump surrounded by a transformed medium mimicks the reflection
of a flat ground, both in electromagnetic and acoustic contexts \cite{pendryprl,pendrynjp}:
Here, the geometric transform is
a one-to-one correspondence between a segment and a curve.
This proposal already led to theoretical \cite{pra2010} and experimental
\cite{ergin} demonstrations of three-dimensional electromagnetic carpets. An experimental
validation of two-dimensional cloaking at visible wavelengths has even been
reported for surface plasmon polaritons at a metal dielectric interface \cite{renger}.
Also woth mentioning is the proposal of a non-singular cloak by
Leonhardt and Tyc based on the
stereographic projection of a hypersphere on a hyperplane, in a way similar to
what is done to design a Maxwell fisheye lens \cite{leonhardt09}.

It is also possible to apply geometric transforms to other wave equations:
However, Milton, Briane, and Willis \cite{milton} have shown
that the elasticity equations are not invariant under coordinate
transformations and consequently that if cloaking
exists for such classes of waves, it would be of a different
nature than its acoustic and electromagnetic counterparts.
A systematic investigation of acoustic cloaking started
with Cummer and Schurig \cite{cummernjp}, who analyzed the two dimensional
cloaking for pressure waves in a transversely
anisotropic fluid by exploiting the analogy with transverse electromagnetic
waves. Chen and Chan \cite{chen07,chenchan2010} and Cummer et al.
\cite{pendryprlb} further noticed that a three-dimensional acoustic
cloaking for pressure waves in a fluid can be envisaged
since the wave equation retains its form under geometric
changes. Farhat et al. also looked at the limitations of
square acoustic cloaks for antiplane shear waves \cite{farhatnjp}.
A promising avenue towards the realization of acoustic
metamaterials was opened by Torrent and Sanchez-Dehesa
\cite{sanchez} and Chen et al. \cite{chen07}, who independently investigated
cloaking for concentric multilayers behaving as anisotropic
fluids in the homogenization limit. Using a similar
approach, Farhat et al. \cite{farhat08} theoretically demonstrated cloaking of
surface liquid waves for a microstructured metallic cloak between 10 Hz and 15 Hz,
which was experimentally validated at 10 Hz.

However, there are other physical situations whereby the mathematical
model for acoustic waves is less tractable. For instance, Norris investigated
some general types of acoustic cloaks with finite mass consisting
of so-called pentamode materials, which display an
anisotropic stiffness \cite{norris}. Also, when one moves to the area of coupled pressure and
shear elastic waves, the isomorphism between the Navier
equations and the scalar wave equation is lost and
computations become more involved. Brun, Guenneau, and Movchan
studied a cylindrical cloak for in-plane elastic waves which
is described by a rank-4 (nonsymmetric) elasticity tensor
with $2^4$ Cartesian entries and an isotropic density \cite{brun}.
Whereas the former structured pentamode metamaterial  \cite{norris} might already
represent a technological challenge for mechanical engineers,
the latter proposal \cite{brun} imposes even severer constraints
on the material parameters. Moreover, the required material
properties for a three-dimensional elastic cloak remain
elusive thus far, as these would involve a rank-4 elasticity
tensor with up to $3^4$ spatially varying nonvanishing Cartesian
entries. However, in the special case of thin-elastic
plates, whose spectral properties require ad hoc numerical
techniques \cite{farhatprb}, it has been shown that the elasticity
tensor can be represented in a cylindrical basis by a diagonal
matrix with two (spatially varying) nonvanishing entries.
It seems therefore quite natural to start designing
such a cloak before investigating the other cases.

In this paper, we investigate the original design of singular cloaks in section 2,
then turn to the analysis of non-singular cloaks
through the blowup of a small ball in section 3 (mimetism) and the blowup of
a line/surface in section 4 (invisibility carpets). We further show in section 5
that an anisotropic cloak can be approximated via a radially symmetric multilayered
cloak filled with piecewise constant isotropic
bulk modulus $\kappa$ and mass density $\rho$ to make an object
surrounded by such a coat neutral for pressure waves in a fluid
background. Finally we consider in section 6 some cloaks with non-smooth boundaries.
Section 7 concludes the paper with some synthesis of our numerical
results and some perspectives for future work.

\section{Transformational acoustics: Spatially varying anisotropic densities and bulk moduli}
\label{chap1}

The key concept in developing an invisibility cloak in electromagnetism and/or
in acoustics is a simple one: it is the
idea that sound and light see space differently. For sound/light, the concept of distance is
modified by the acoustic/electromagnetic properties of the regions through which it
travels. In geometrical acoustics/optics, we are used to the idea of the acoustical/optical
path; when travelling an infinitesimal distance $ds$, the corresponding acoustical/optical
path length is $n ds$, where $n$ is the local refractive index, equal to
$\sqrt{\varepsilon\mu}$ in the electromagnetic setting (with $\varepsilon$ the permittivity
and $\mu$ the permeability of the medium), and equal to $\sqrt{\rho/\kappa}$ (with $\rho$ the fluid
density and $\kappa$ its compressional modulus). This is the first hint
that it may be possible to mimic a transformation of a region of space by using
an equivalent transformation of acoustic/electromagnetic properties. We report in Figure \ref{cloaking-fig0}
some finite element computations for the scattering of a pressure wave by a rigid obstacle on its own (lower
panel) and when it is surrounded by a singular cloak (upper panel) obtained from blowing up a point in a ball.

\begin{figure}[h!]
\centering
\includegraphics[scale=0.60]{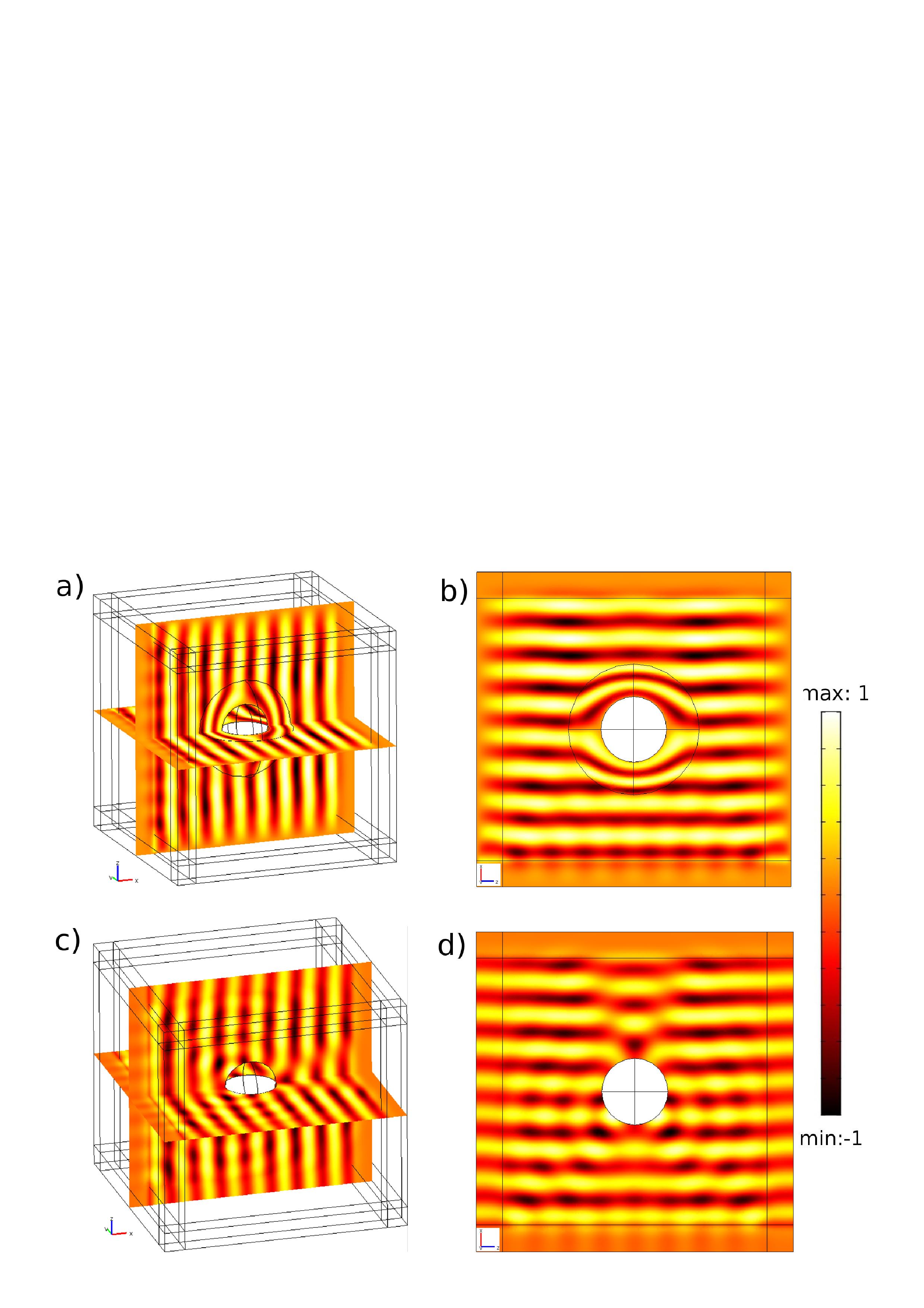}
\vspace{0cm}
\caption{\footnotesize~l Real part of the pressure field $p$ scattered by
a rigid obstacle (Neumann boundary conditions) of radius $0.5$m  on its own (c,d) and when it
is surrounded by a cloak of inner
radius $R_1=0.5$ m, outer radius $R_2=1$ m (a,b) for an incoming plane
wave of wavelength $\lambda=0.3$m. (a,c) Three-dimensional plots,
the pressure wave is propagating along the x direction (from left to right); (b,d)
Two-dimensional plots in the (zx)-plane, the pressure wave
propagates along the vertical x direction (from bottom to top).
We note the strong backward scattering in panels (c) and (d).}
\label{cloaking-fig0}
\end{figure}

\subsection{Change of coordinates: Gradient behaviour.}

If one considers the geometric transform from the coordinate system $(x,y,z)$ to $(u,v,w)$,
the relation between the gradient operator in both coordinate systems is given by:\\

\begin{equation}
 \left( \begin{array}{c}
 \dfrac{\partial}{\partial x}\\ \\
\dfrac{\partial}{\partial y}\\ \\
\dfrac{\partial}{\partial z}\\
\end{array} \right) =
 \left( \begin{array}{c c c}
 \dfrac{\partial u}{\partial x} & \dfrac{\partial v}{\partial x} & \dfrac{\partial w}{\partial x}\\ \\
\dfrac{\partial u}{\partial y} & \dfrac{\partial v}{\partial y} & \dfrac{\partial w}{\partial y}\\ \\
\dfrac{\partial u}{\partial z} & \dfrac{\partial v}{\partial z} & \dfrac{\partial w}{\partial z}
\end{array} \right) \,
 \left( \begin{array}{c}
 \dfrac{\partial}{\partial u}\\ \\
\dfrac{\partial}{\partial v}\\ \\
\dfrac{\partial}{\partial w}\\
\end{array} \right) {\bf\Longleftrightarrow} \nabla_{(x,y,z)} = \left( {\bf J}_{ux} \right)^T.\nabla_{(u,v,w)} \label{grad}
\end{equation}\\

where ${\bf J}_{ux} = \left( {\bf J}_{xu} \right)^{-1}=\partial(u,v,w)/\partial(x,y,z)$ is the Jacobian of the transformation
$(u,v,w) \rightarrow (x,y,z)$.\\

\subsection{Transformational acoustics and weak formulation of the wave equation}

We now invoke the three-dimensional acoustic equation: \\

\begin{equation}
\nabla . \left( \underline{\underline{\rho}}^{-1} \nabla p \right) + \frac{\omega^2}{\kappa}p = 0\, ,\qquad
\hbox{in }\,\Omega\, ,
\label{equ_ac}
\end{equation}
where $\Omega$ is a bounded domain of $\mathcal{R}^3$.

Multiplying (\textit{\ref{equ_ac}}) by a test function $\bar{\phi}$ and integrating by parts, we obtain:\\

\begin{equation}
 -\int_{\Omega}dV \left( \nabla_{(x,y,z)} \bar{\phi} . \underline{\underline{\rho}}^{-1} \nabla_{(x,y,z)} p \right) +
\int_{\Omega}dV \left( \omega^2 \kappa^{-1} p \bar{\phi} \right) = 0
\label{equ_ac_weak}
\end{equation}
where we have assumed that the surfacic term is zero (corresponding to a Neumann integral over the boundary
$\partial\Omega$).

We now apply to equation (\textit{\ref{equ_ac_weak}}) the coordinate change $(x,y,z) \rightarrow (u,v,w)$
and using (\textit{\ref{grad}}), we deduce:\\

\begin{equation}
 -\int_{\Omega}dV' \left\{  \left( {\bf J}_{ux}^T \nabla_{(u,v,w)} \bar{\phi} . \underline{\underline{\rho}}^{-1}{\bf J}_{ux}^T \nabla_{(u,v,w)} p \right) det({\bf J}_{xu})\right\}
+\int_{\Omega}dV' \left( det({\bf J}_{xu}) \omega^2 \kappa^{-1} p \bar{\phi} \right) = 0
\end{equation}\\

And this can be expressed in terms of ${\bf J}_{ux}$:\\

\begin{equation}
 -\int_{\Omega}dV' \left(  \left( \nabla_{(u,v,w)} \bar{\phi} \right)^T \dfrac{{\bf J}_{ux}.\underline{\underline{\rho}}^{-1}.{\bf J}_{ux}^T}{det({\bf J}_{ux})} \nabla_{(u,v,w)} p \right)
+\int_{\Omega}dV' \left( \dfrac{\kappa^{-1}}{det({\bf J}_{ux})} \omega^2  p \bar{\phi} \right) = 0
\end{equation}\\

The new parameters $\underline{\underline{\tilde{\rho}}}$ and $\tilde{\kappa}$ after the change of coordinates can be now deduced, and are given by:\\

\begin{equation}
\left\{ \begin{array}{c c l c r}
 \underline{\underline{\tilde{\rho}}} & = & {\bf J}_{ux}^{-T}.\underline{\underline{\rho}}.{\bf J}_{ux}^{-1}.det({\bf J}_{ux})  &  & \\
 & & & & \text{in terms of } {\bf J}_{ux}  \\
\tilde{\kappa} & = & \kappa det({\bf J}_{ux}) & &
\end{array} \right.
\hbox{ or }
\left\{ \begin{array}{c c l c r}
 \underline{\underline{\tilde{\rho}}} & = & \dfrac{{\bf J}_{xu}^{T}.\underline{\underline{\rho}}.{\bf J}_{ux}}{det({\bf J}_{xu})}  &  & \\
 & & & & \qquad \text{\; \; \; in terms of } {\bf J}_{xu}  \\
\tilde{\kappa} & = & \dfrac{\kappa}{det({\bf J}_{xu})} & &
\end{array} \right.
\end{equation}

\section{Blow-up of a small ball: Non-singular cloaking and anamorphism}
\label{chap3}
In 2008, Kohn and co-authors \cite{kohn} have regularized the transform of Pendry et al. which blows up a point
onto a sphere \cite{pendry}.
The former authors consider a small ball of radius $R_0$ and define its blow up into
a unit ball  in the following way:

\begin{itemize}

 \item $0 \leq r \leq R_0, ~~~~ ~~~~ ~~
\left\{
\begin{array}{l c l}
  r' & = & \dfrac{R_1}{R_0} r \\
  \theta' & = & \theta \\
  \phi' & = & \phi
\end{array}
\right.
$
 \item $R_0 \leq r \leq R_2, $ ~~~~~~~~~~ 
$
\left\{
\begin{array}{l c l}
  r' & = & \dfrac{R_2 - R_1}{R_2 - R_0} r + R_2 \dfrac{R_1 - R_0}{R_2 - R_0}\\
  \theta' & = & \theta \\
  \phi' & = & \phi
\end{array}
\right.
$
\end{itemize}
Importantly, this transform is continuous and piecewise smooth and this transform reduces
to the identity on the boundary $r=R_2$. This latter property ensures that the cloak will
be impedanced matched to the surrounding medium, in order to avoid any reflection on the
cloak outer boundary.  Diatta et al. realized that such a non-singular approach to cloaking
leads to anamorphism \cite{diatta2010,diatta2011}, whereby a rigid
object scatters pressure waves like another rigid object (of different size and/or shape).

The Jacobian of the transformation is given by :\\

${\bf J}_{rr'} = \left(
\begin{array}{c c c}
  \alpha^{-1} & 0 & 0 \\
  0 & 1 & 0\\
  0 & 0 & 1 \\
\end{array} \right) \quad \text{with} \quad
\left\{
\begin{array}{l c l l}
  \alpha & = & \dfrac{R_1}{R_0} & (r \leq R_0) \\
  & & & \\
  \alpha & = & \dfrac{R_2 - R_1}{R_2 - R_0} & (R_0 \leq r \leq R_2) \\
\end{array}
\right.
$\\

\begin{figure}[h!]
\centering
\includegraphics[scale=.60]{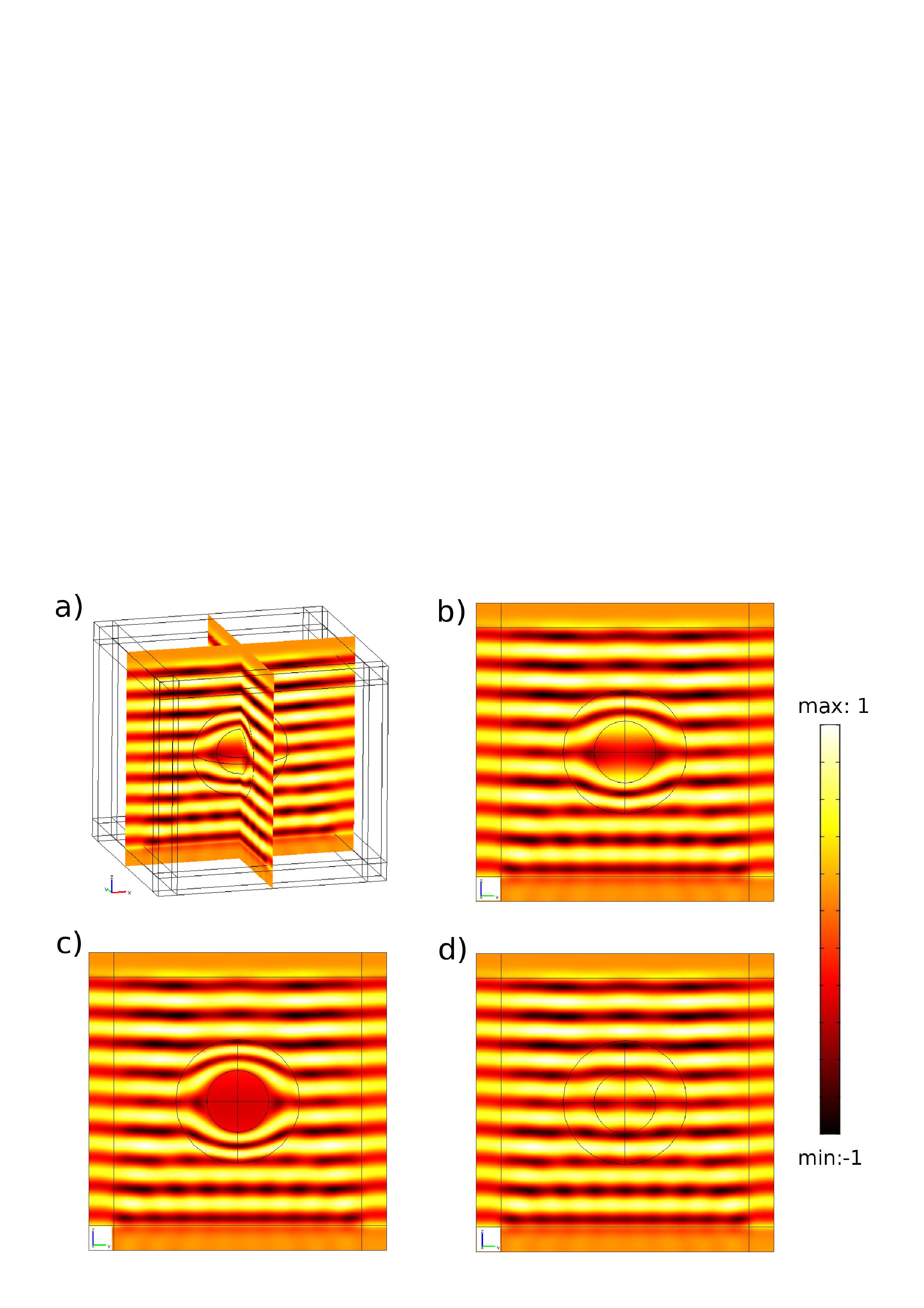}
\caption{ Real part of the pressure field $p$ scattered by
a non-singular cloak of inner radius $0.5$m and outer radius $1$m for an incoming plane
wave of wavelength $\lambda=0.45$m propagating along the vertical z-direction
(from bottom to top);
(a) Three-dimensional view for $r_0=0.2$m; (b) 2D view in the yz-plane for $r_0=0.2$m;
(c) 2D view in the yz-plane for $r_0=0.05$m; (d) 2D view in the yz-plane for $r_0=0.45$m;
We note that the larger $r_0$, the smaller the scattering. In the limit case $r_0=0.5$m,
the cloak is filled with homogeneous isotropic fluid.}
\label{kohn-fig}
\end{figure}

Thus, the transformation matrix for the two media is given by

${\bf T}^{-1} =
\left(
\begin{array}{c c c}
  \dfrac{(r - \beta)^2}{\alpha r^2} & 0 & 0 \\
  0 & \alpha^{-1} & 0 \\
  0 & 0 & \alpha^{-1} \\
\end{array}
\right) \quad \text{with} \quad
\left\{
\begin{array}{l c l l}
  \alpha & = & \dfrac{R_1}{R_0} \quad ; \quad \beta = 0 & (r \leq R_0) \\
  & & & \\
  \alpha & = & \dfrac{R_2 - R_1}{R_2 - R_0} \quad ; \quad \beta = R_2 \dfrac{R_1 - R_0}{R_2 - R_0} & (R_0 \leq r \leq R_2) \\
\end{array}
\right.
$\\

We note that ${\bf T}^{-1}$ has two constant non-zero eigenvalues, and one spatially varying eigenvalue which tends to
$\alpha^{-1}=R_1/R_0$ when $r$ goes to zero. Such an invisibility cloak, which is made of two heterogeneous
anistropic media (inside the ball $r\leq R_1$ and inside the shell $R_1<r<R_2$) is therefore non singular.
We report in Figure \ref{kohn-fig} some finite element computations showing that when the cloak is designed from a ball
with very small radius (c), there is no field inside the inner region of radius $R_1$ (nearly a singular cloak), while for
a radius $R_0\sim R_1$ the cloak and the object inside it do not curve the wavefront of the incident wave (d): in that case
the anisotry is very small.
The case $R_0 = 0$ corresponds to a cloak with a heterogeneous anisotropic shell $R_1<r<R_2$, and some vaccuum
inside the ball $r\leq R_1$. In this way, the domain $r < R_1$ is now an invisible domain for all exterior observator.
However, we note that in this case, the first eigenvalue tends to infinity when $r$ tends to $R_1$
so that ${\bf T}^{-1}$ now has two constant eigenvalues and one which tends to infinity on its inner boundary $r=R_1$.
One of two things can thus happen: one either considers a regular cloak with an invisibility region filled with metamaterial;
or one considers a singular cloak with an invisibility region filled with vaccuum. These two cases have their own
advantages and downsides: The former leaves no room for a object to hide, whereas the latter can not be achieved in
practice.




\section{Blow-up of a line: Carpets for curved grounds}
\label{chap2}


Another type of cloak is discussed in this section: one that gives all cloaked objects the appearance of
a flat conducting sheet in the context of electromagnetism \cite{pendryprl,ergin,pra2010},
or a flat rigid ground in the context of acoustics \cite{pendrynjp}.
It has the advantage that none of the parameters of the cloak
is singular and can in fact be made isotropic using quasi-conformal mappings.
It makes broadband cloaking in the optical frequencies possible, as chiefly demonstrated for surface
plasmon polaritons at a wavelength of 800 nanometers (requiring a specific arrangement of
about 100 dielectric pillars 200 nanometers in diameter) \cite{renger}.
This alternative approach to invisibility cloaks takes advantage of the principle that a perfect
conducting sheet is inherently invisible to a specifically polarized incidence when the electric field is
perpendicular to the conducing sheet. Then if there is a bump on
the ground, invisibility can be realized by guiding the incident wave around the bump,
keeping the electric field always perpendicular to the curved surface, which is the
function of the carpet cloak proposed by Li and Pendry.

Let us consider the linear geometric transform :
\begin{equation}
x' = x, ~~~
y' = y, ~~~
z' = \frac{z_2(x,y)-z_1(x,y)}{z_2(x,y)}z + z_1(x,y).
\end{equation}

The linear transform is expressed in a Cartesian basis as:
${\bf J}_{xx'}=\left(
\begin{array}{ccc}
                1 & 0 & 0\\
             0 & 1 & 0\\
             \frac{\partial z}{\partial x'} & \frac{\partial z}{\partial y'} & \frac{1}{\alpha}\\
               \end{array}\right)$
where $\alpha=\dfrac{z_2(x,y)-z_1(x,y)}{z_2(x,y)}$ and from the chain rule

\begin{equation}
\frac{\partial z}{\partial x'}= z_2 \frac{z'-z_2}{{(z_2-z_1)}^2}
\frac{\partial z_1}{\partial x}- z_1 \frac{z'-z_1}{{(z_2-z_1)}^2}
\frac{\partial z_2}{\partial x} \; , \; \frac{\partial z}{\partial
y'}= z_2 \frac{z'-z_2}{{(z_2-z_1)}^2} \frac{\partial z_1}{\partial
y}- z_1 \frac{z'-z_1}{{(z_2-z_1)}^2} \frac{\partial z_2}{\partial
y}\; .
\end{equation}

This leads to the inverse symmetric tensor ${\bf T}^{-1}$ which is fully described by
seven non vanishing entries in a Cartesian basis:

\begin{center}
 \begin{tabular}{ll}
$(T^{-1})_{11}=(T^{-1})_{22}\displaystyle \frac{1}{\alpha}$ &,
$(T^{(-1)})_{13}=(T^{-1})_{31}=-\displaystyle \frac{\partial
z}{\partial x'}$\\
$(T^{-1})_{23}=(T^{-1})_{32}=-\displaystyle \frac{\partial
z}{\partial y'}$ &, $(T^{-1})_{33}=\displaystyle \left(
1+{\left(\frac{\partial z}{\partial
x'}\right)}^2+{\left(\frac{\partial z}{\partial y'}\right)}^2\right)
\alpha$
\end{tabular}
\end{center}

We report some finite element computations for a carpet with parabolic shape in Figure \ref{carpet-fig}. A pressure
wave incident from above on a flat ground with a parabolic bump (lower panel) is backscattered with a disturbed
wavefront, while the wavefront is flattened when the bump is surrounded by a carpet (upper panel).

\begin{figure}[h!]
\centering
\includegraphics[scale=0.60]{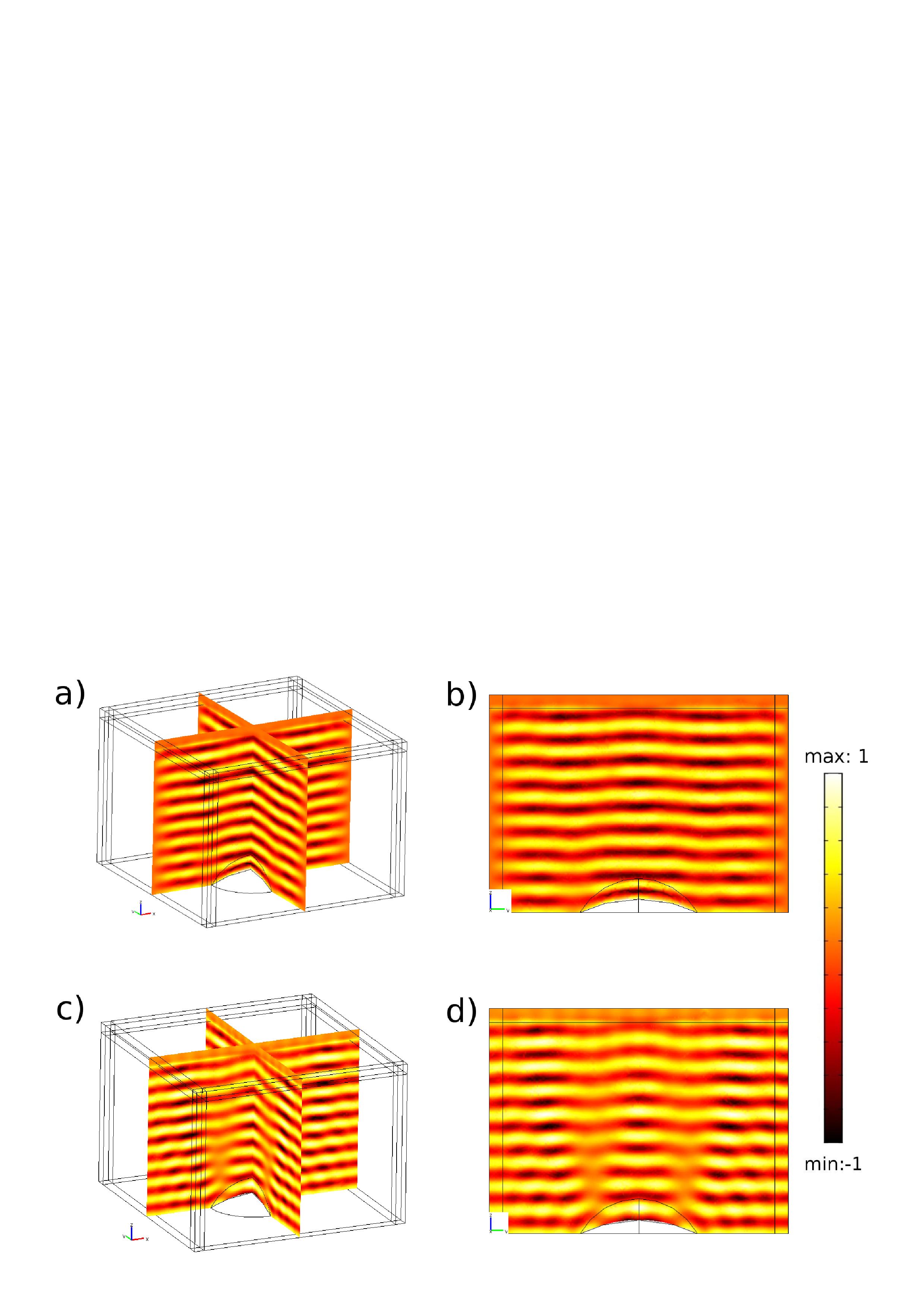}
\vspace{0cm}
\caption{\footnotesize Lower panel: Pressure wave of wavelength $\lambda=0.3$m incident from above on a flat (rigid) ground with a bump described by the surface
$z_1(x,y) = \sqrt{4-x^2-y^2}-1.802776$;
Upper panel: Pressure wave of wavelength $\lambda=0.3$m incident from above on a carpet of parabolic shape with inner and outer boundaries
$z_1(x,y))$ and $z_2(x,y) = \sqrt{1-x^2-y^2}-1/2$ surrounding the same bump as aforementioned; (a) and (b) are three-dimensional plots, while (c) and (d)
are two-dimensional plots along the vertical slices.}
\label{carpet-fig}
\end{figure}

\section{Approximating ideal cloaks: Structured cloaks}
\label{chap4}


Transformational acoustics is a nice way to imagine new kinds of metafluids, but it is
however important to be able to fabricate these with materials/fluids at hand. To
achieve this practical goal, one possible route is to use homogenization techniques
which have been developed over the past 40 years by applied mathematicians \cite{bens,jko94}.
Such multiscale techniques make effective medium theories rigorous and further allow for more
flexibility in the geometrical and material parameters: one can homogenized materials with
small inclusions of any shapes  in two or three space dimensions,
and these can be anisotropic.

\subsection{Homogenization of the acoustic equation}
In our case, the pressure field is solution of the following governing equation:
\begin{equation}
\nabla\cdot\left(\lambda_{\eta}\nabla p\right)+\dfrac{\omega^2}{c^2}\zeta_{\eta}p=0\, ,
\label{acoustic-hom1}
\end{equation}
inside the heterogeneous isotropic cloak $\Omega_f$, where
$$\zeta_\eta=\kappa^{-1}(\frac{r}{\eta}) \; \hbox{ and }
\; \lambda_\eta=\rho^{-1}(\frac{r}{\eta}) \; .$$

Furthermore, $c^2=\kappa_0/\rho_0$, where $\kappa_0$ is the bulk modulus of
the fluid and $\rho_0$ its density.

When the acoustic wave penetrates the structured cloak $\Omega_f$, it
undergoes fast periodic oscillations. To filter these oscillations,
we consider an asymptotic expansion of the associated pressure $p$
solution of (\ref{acoustic-hom1}) in terms of a
macroscopic (or slow) variable ${\bf x}=(r,\theta)$ and a
microscopic (or fast) variable ${\bf
x}_\eta=(\frac{r}{\eta},\theta)$, where $\eta$ is a small positive
real parameter.

The homogenization of this elliptic equation can be derived by considering
the following Ansatz
\begin{equation}
\forall {\bf x} \in \Omega_f, \;\; p_\eta ({\bf x}) =
\sum_{i=0}^\infty \eta^i p^{(i)}({\bf x},{\bf x}_\eta) \; ,
\end{equation}
where $\forall {\bf x}\in\Omega_f,\; p^{(i)}({\bf x},\cdot)$ is
1-periodic along $r$. Note that we evenly divide $\Omega_f$
($R_1\leq r \leq R_2$, $0\leq\theta<2\pi$) into a large number of
thin curved layers of radial length $(R_2-R_1)/\eta$, but the
spectral parameter $\beta_0$ in Eq. (\ref{acoustic-hom1}) remains fixed
(so is the wave frequency).

Rescaling the differential operator in Eq. (\ref{acoustic-hom1})
accordingly as $\nabla=\nabla_{\bf x}+{\bf
e}_r\displaystyle{\frac{1}{\eta}\frac{\partial}{\partial r}}$, and
collecting the terms sitting in front of the same powers of $\eta$
(see e.g. \cite{bens,jko94}), we obtain
\begin{equation}
\nabla\cdot\left(\underline{\underline{\lambda}}\nabla p\right)+\dfrac{\omega^2}{c^2}<\zeta>p=0 \; ,
\label{acoustic-hom2}
\end{equation}
which is the homogenized acoustic equation where
$<\zeta>=\int_0^1\kappa^{-1}(r)\, dr$ and with
$\underline{\underline{\lambda}}$ a homogenized rank-2 diagonal tensor
which has the physical dimensions of a homogenized anisotropic density
$\underline{\underline{\rho}}={\rm Diag}(\rho_r,\rho_\theta,\rho_\phi)$ given by
\begin{equation}
\underline{\underline{\lambda}}={\rm
Diag}({<\lambda^{-1}>}^{-1},<\lambda>,<\lambda>)=\underline{\underline{\rho}}^{-1}
\; .
\label{parameter2}
\end{equation}

We note that if the cloak consists of an alternation of two
homogeneous isotropic layers of thicknesses $d_A$ and $d_B$ and
bulk moduli $\kappa_A$, $\kappa_B$ and densities $\rho_A$ and $\rho_B$, we
have
\begin{equation}
\begin{array}{lll}
&\displaystyle{\frac{1}{\rho_r}}=\displaystyle{\frac{1}{1+\eta}\left(\frac{1}{\rho_A}+\frac{\eta}{\rho_B}\right)},
&\rho_\theta=\rho_\phi=\displaystyle{\frac{\rho_A+\eta \rho_B}{1+\eta}} \; , \;
~~ \kappa=\displaystyle{\frac{\kappa_A+\eta \kappa_B}{1+\eta}} \; ,\nonumber
\end{array}
\end{equation}
where $\eta=d_B/d_A$ is the ratio of thicknesses for layers $A$ and
$B$ and $d_A+d_B=1$.

We now note that the coordinate transformation
$r'=R_1+r\frac{R_2-R_1}{R_2}$ can compress the region $r<R_2$ into
the shell $R_1<r<R_2$, provided that the cloak is described by the
following density and bulk modulus
\begin{equation}
\begin{array}{lll}
\rho_r &=\displaystyle{\frac{R_2-R_1}{R_2}{\left(\frac{r}{r-R_1}\right)}^2} \, , \; ~~~
\rho_{\theta}=\rho_\phi=\displaystyle{\left(\frac{R_2-R_1}{R_2}\right)} \, ,
~~~ \kappa &
=\displaystyle{{\left(\frac{R_2-R_1}{R_2}\right)}^3\,{\left(\frac{r}{r-R_1}\right)}^2}
\; ,
\end{array}
\label{rhort1}
\end{equation}
where $R_1$ and $R_2$ are the interior and the exterior radii of the
elastic coat of thickness $h$.

To mimic these fluid parameters, we proceed in two steps,
following \cite{chen07}: we first approximate the ideal cloak by a
multi-layered cloak with $M$ anisotropic homogeneous concentric
layers. We then approximate each layer $i$, $i=1,.., M$ by $N$ thin
isotropic layers through the homogenization process described above.
This means the overall number $NM$ of isotropic layers can be fairly
large.

\subsection{Numerical results for a multi-layered cloaking device}

\begin{figure}[h]
\centering
\includegraphics[scale=0.60]{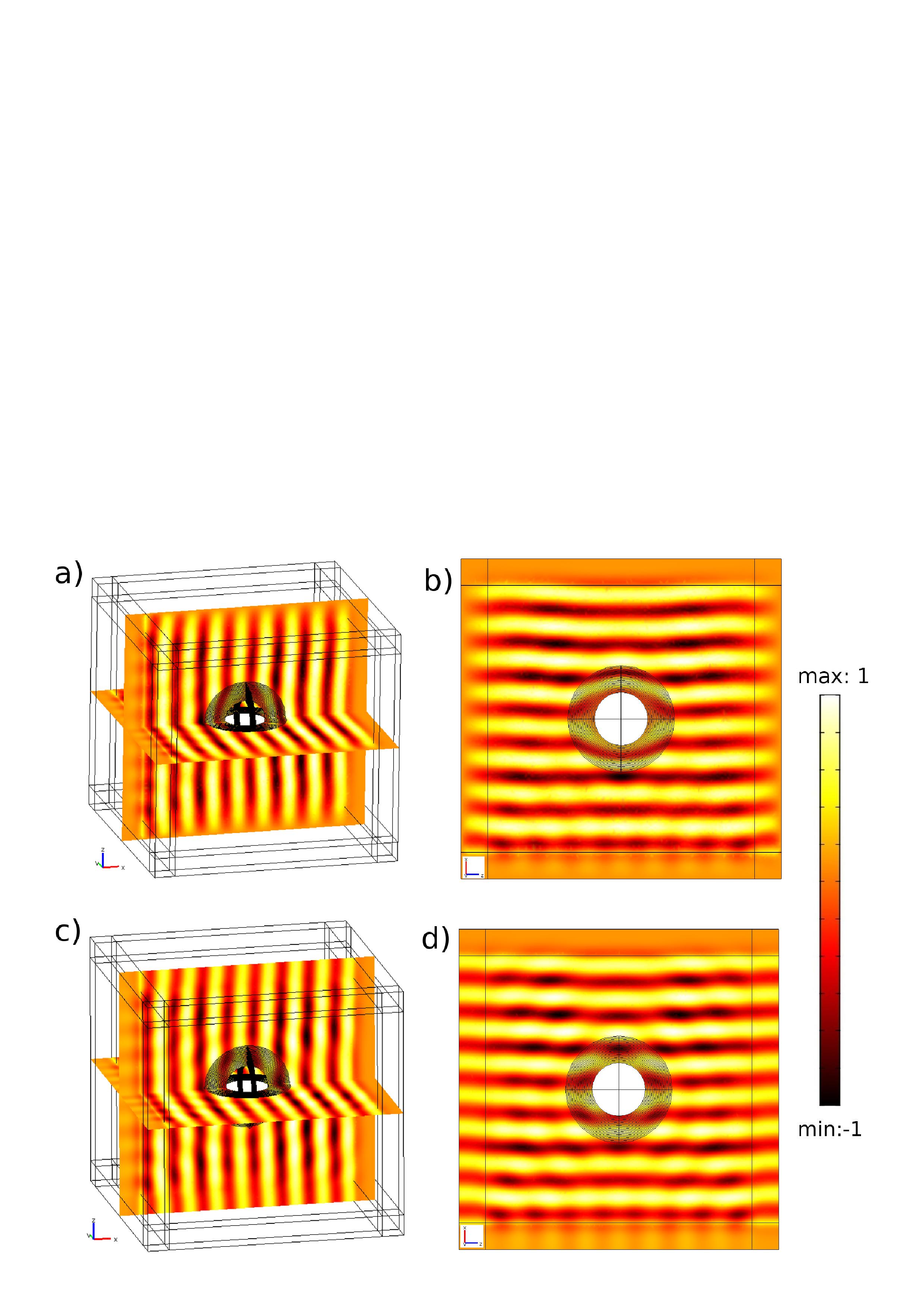}
\vspace{0cm}
\caption{\footnotesize ~ Real part of the pressure field $p$ scattered by
a rigid obstacle of radius $R_1=0.2$m surrounded by a non-singular multi-layered
cloak of inner radius $R_1$ and outer radius $R_2=0.4$m for an incoming plane
wave of wavelength $\lambda=0.25$m. These cloaks scatter waves like
a rigid obstacle of radius $r_0$ (mimetism). Upper panel:
Cloak scattering like a rigid sphere of radius $r_o = 0.05m$ , the ranges of parameters are $\rho_i/\rho_o = [0.0243, 3.4636]$ and $\kappa_i/\kappa_o = 0.5714$;
Lower panel: Cloak scattering like a rigid sphere of radius $r_o = 0.15$m, with ranges of parameters $\rho_i/\rho_o = [0.2601,2.2229]$;
$\kappa_i/\kappa_o = 0.8$. We note that the density takes less extreme values for $r_0=0.15$m.
(a,c) Three-dimensional plot (the pressure wave is incident from the left);
(b,d) Two dimensional plot in the vertical $zx$-plane (the pressure wave is incident from below).}
\label{multilayer}
\end{figure}

\begin{figure}[h]
\centering
\includegraphics[scale=0.60]{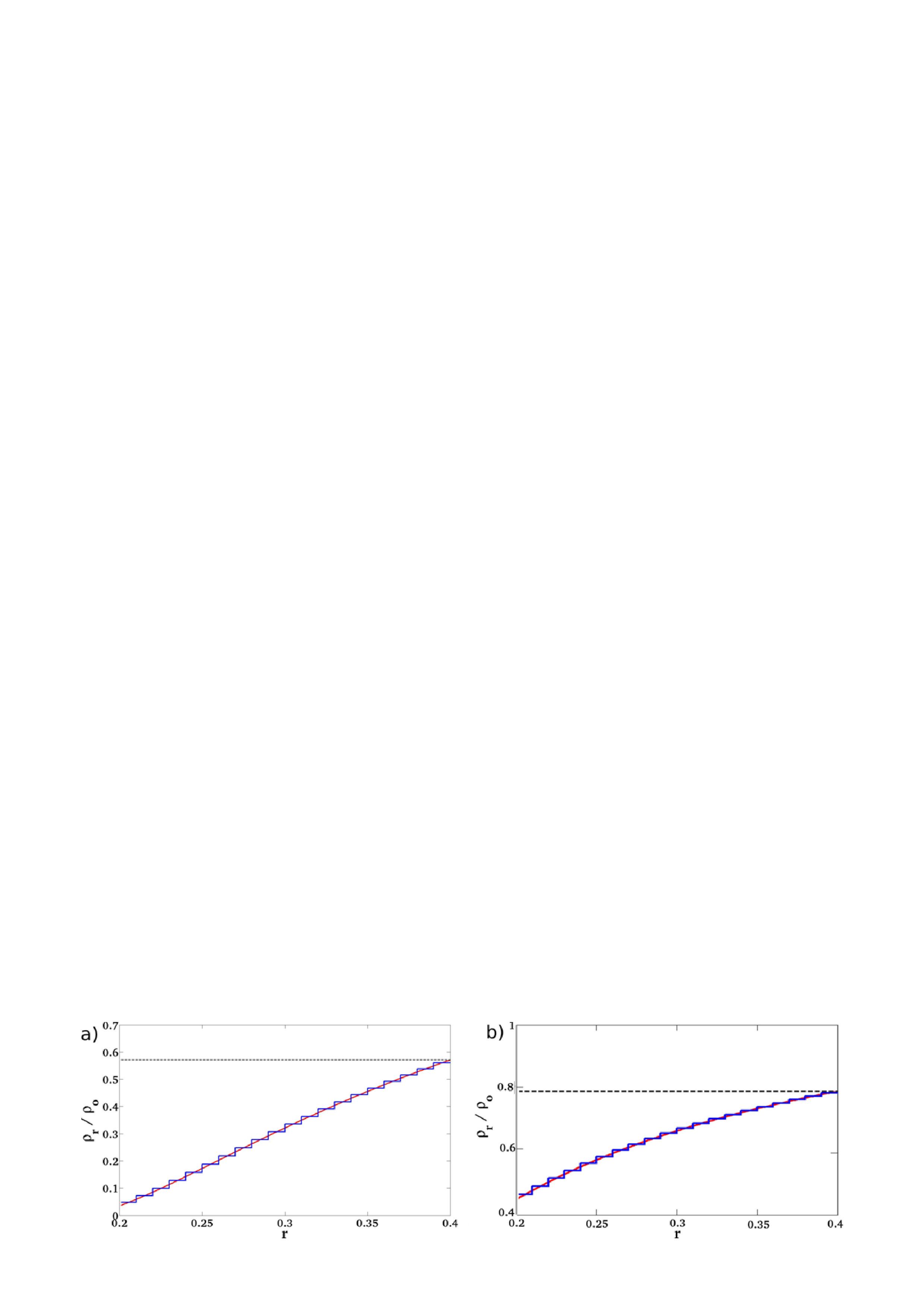}
\vspace{0cm}
\caption{\footnotesize ~ Variation profil of the reduced parameter $\rho_{r}/\rho_{0}$ versus the thikness of the cloak
(red curve) and a layered approach (blue staircase curve). Here, we consider the blowup of a small
ball of radius $r_0=0.05$m (a) and a small ball of radius $0.15$m (b). Importantly, the lower bound
for the density in (b) is about five times larger, while the upper bound is similar to that in (a).
It should be also noted that the graph in
(b) is the restriction of the graph in Fig. \ref{kohn_m} to the interval $r\in[0.2m,0.4m]$.}
\label{nokohn}
\end{figure}

\begin{figure}[h!]
\centering
\includegraphics[scale=0.60]{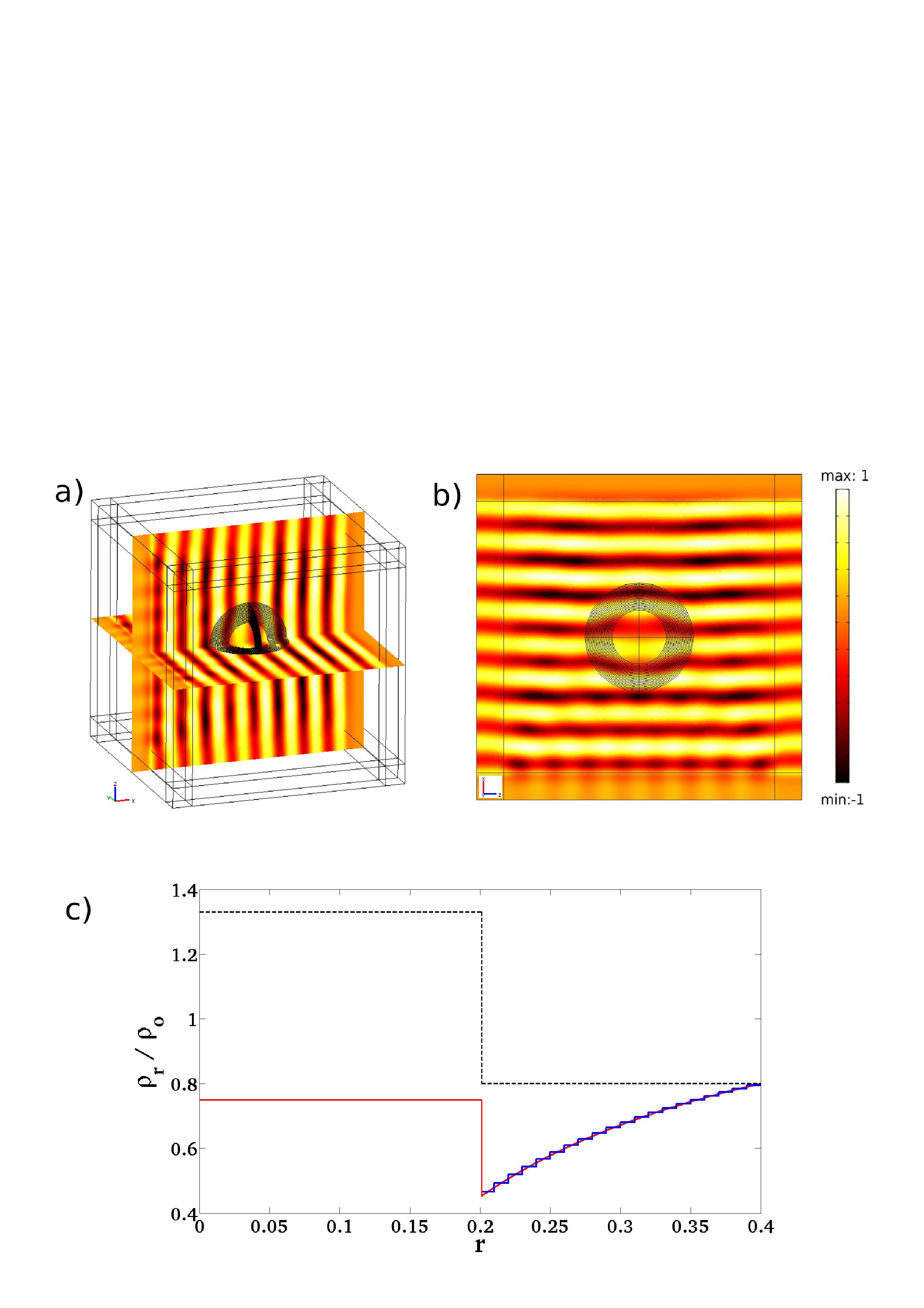}
\vspace{0cm}
\caption{\footnotesize  Real part of the pressure field $p$ scattered by
a non-singular multi-layered cloak of inner radius $R_1=0.2$m and outer radius $R_2=0.4$m for an incoming plane
wave of wavelength $\lambda=0.25$m. We consider the blowup of a small ball of radius $r_0=0.15$m.
(a) Three-dimensional plot (the pressure wave is incident from the left);
(b) Two dimensional plot in the $zx$-plane (the pressure wave is incident from below);
(c) The variation of the bulk modulus versus the radiusq
is given by the dashed black curve,
while the variation of the density (solid red curve) is approximated by the blue stair-case curve. Importantly,
the cloak is filled with a homogeneous isotropic fluid of density
$\rho_r/\rho_o = 0.75$ and bulk modulus $\kappa_r/\kappa_o  = 1.333$.
Importantly, the density and bulk modulus display a jump at the inner boundary of the cloak $R_1$ and they
are both constant in the core i.e. for $r<R_1$ (isotropic homogeneous fluid). It is therefore counter-intuitive
that the system core and cloak is invisible (while the core and cloak taken separately are not).}
\label{kohn_m}
\end{figure}

To check the functionality of the multi-layered acoustic cloak, we have performed
some finite element simulations.
Maps of the pressure wave at wavelength $0.25$m propagating at a time instant are
represented by the real part
of the complex amplitude $p$ from the left in Fig. \ref{multilayer}
and is scattered by a rigid circular
obstacle (left panel a) and by the layered cloak (right panel
b). When it is surrounded by the  heterogeneous cloak, which
consists of a circular coating of radii $R_1=0.2$ and $R_2=0.4$ meters,
both forward and backward scattering nearly vanish, with
interestingly a phase shift between the acoustic wave propagating in
homogeneous fluid and the wave bent by the cloak (the
acoustic path is obviously much different).
We note that this phase shift cannot be avoided in electromagnetic metamaterial
cloaks, as otherwise the wave propagating within the cloak
would have to travel faster than the speed of light in vacuum. The acoustic parameters
of the proposed layered cloak are characterized by a scalar bulk modulus $\kappa$
and a spatially varying rank $2$ tensor $\underline{\underline{\rho}}$ given
by (\ref{rhort1}), see Figure \ref{multilayer} for the plots of the pressure field
scattered by such a cloak and Figure \ref{nokohn} for the parameters of the cloak.
Moreover,  we numerically checked that the cloak is broadband in nature.


To further demonstrate the validity of our approach to broadband
cloaking, we now propose an original design for the non-singular
cloak proposed by Kohn et al. \cite{kohn} (i.e. deduced from
the blow up of a small ball instead of a point) with $20$
concentric layers of homogeneous fluids, see Fig. \ref{kohn_m}.
We note that the constraint on the fluid parameters is relaxed
(the lower bound for the density is higher). We hope this simplified set
of material parameters will pave the way towards an experimental
realization of a near-ideal acoustic cloak in the near future.

\vfill\eject

\section{Mimesis with Polyhedron and Star}
One of the specific features of polyhedra and stars is their symmetries, allowing to split the cloak into sub-regions with (smooth) linear boundaries.
Such  sub-regions are diffeomorphic via a linear action of elements of a symmetry group. This allows to simplify the design of such cloaks as
we study one fixed sub-domain, and deduce the properties of the other remaining ones using the diffeomorphisms mapping it to the other ones, which are parts of a symmetry group acting on the whole domain. For the two-dimensional case, we refer the reader to \cite{diatta2009}.

\subsection{Design of three-dimensional cloaks with corners}

For this construction, we use spherical coordinates $(\rho,\theta,\phi)$ related to cartesian coordinates by
\begin{eqnarray}
\left\{
\begin{array}{lr}
\rho=\sqrt{x^2+y^2+z^3}, ~~~~\\
\theta=\arctan(y/x),~~~~\\
\phi=\arccos(\frac{z}{\sqrt{x^2+y^2+z^2}})\\
\end{array}
\right.
\end{eqnarray}
where $\rho,\theta$ and $\phi$ belong to the appropriate subsets of $\mathbb R$ in accordance to the geometry of the cloak. Or equivalently, we have
\begin{eqnarray}
\left\{
\begin{array}{lr}
x=\rho\cos(\theta)\sin(\phi), ~~~~\\
y=\rho\sin(\theta)\sin(\phi), ~~~~\\
z=\rho\cos(\phi)
\end{array}
\right.
\end{eqnarray}
 We denote by ${\bf J}_{x\rho}$ the Jacobian of the latter change of coordinates. That is,
\begin{eqnarray}{\bf J}_{x\rho}={\bf J}_{\phi}diag(1,\rho,\rho)\text{ with } {\bf J}_{\phi} :=\begin{pmatrix}\cos(\theta)\sin(\phi)& -\sin(\theta)\sin(\phi) & \cos(\theta)\cos(\phi)\\
\sin(\theta)\sin(\phi)& \cos(\theta)\sin(\phi) & \sin(\theta)\cos(\phi)\\
\cos(\phi)& 0 & -\sin(\phi)
\end{pmatrix} \end{eqnarray}
To express the transformation $(\rho,\theta,\phi)\mapsto (\rho'(\rho,\theta,\phi), \theta'(\rho,\theta,\phi),\phi'(\rho,\theta,\phi)),$ we consider three noncollinear points $(x_{i,1},y_{i,1},z_{i,1}), ~ (x_{i,2},y_{i,2},z_{i,2}), (x_{i,3},y_{i,3},z_{i,3}),$ on each $S_i$  of the boundary surfaces $S_0, S_1,S_2$ of the sub-domain, then we use
$$
\left\{
\begin{array}{lr}
\rho'= R_1+\frac{R_2-R_1}{R_2-R_0}(\rho-R_0), ~~~~\\
\theta'=\theta, ~~~~\\
\phi'=\phi
\end{array}
\right.
$$
where $R_i$ are now smooth functions $R(\theta,\phi)$ of $\theta$ and $\phi)$
\begin{eqnarray} R_i&=&\frac{d_i}{a_i\cos(\theta)\sin(\phi)+b_i\sin(\theta)\sin(\phi) + c_i\cos(\phi)}, i=0,1,2\end{eqnarray}
 and $a_i, ~b_i, ~c_i, ~d_i$ are constants
 \begin{eqnarray}a_i& =&(y_{i,2}-y_{i,1})(z_{i,3}-z_{i,1})-(z_{i,2}-z_{i,1})(y_{i,3}-y_{i,1}),\nonumber\\
b_i &=&(z_{i,2}-z_{i,1})(x_{i,3}-x_{i,1})-(x_{i,2}-x_{i,1})(z_{i,3}-z_{i,1}), ~ ~
 c_i~ =~ (x_{i,2}-x_{i,1})(y_{i,3}-y_{i,1})-(y_{i,2}-y_{i,1})(x_{i,3}-x_{i,1}),\nonumber\\
 d_i
 &=& (y_{i,2}z_{i,3}-y_{i,2}z_{i,1}-y_{i,1}z_{i,3}-z_{i,2}y_{i,3}+z_{i,2}y_{i,1}+z_{i,1}y_{i,3})
 x_{i,1}+(z_{i,2}x_{i,3}-z_{i,2}x_{i,1}-z_{i,1}x_{i,3}-x_{i,2}z_{i,3}+x_{i,2}z_{i,1} \nonumber\\
 &+& x_{i,1}z_{i,3})y_{i,1}
 + (x_{i,2}y_{i,3}-x_{i,2}y_{i,1}-x_{i,1}y_{i,3}-y_{i,2}x_{i,3}+y_{i,2}x_{i,1}+y_{i,1}x_{i,3})z_
{i,1}.\end{eqnarray}

Now if we set $\alpha:=\frac{R_2-R_1}{R_2-R_0},$
\begin{eqnarray}R_{i,\theta}&:=& \frac{\partial R_i}{\partial\theta} = \frac{d_i(a_i\sin(\theta)\sin(\phi)-b_i\cos(\theta)\sin(\phi))}{(a_i\cos(\theta)\sin(\phi)+b_i\sin(\theta)\sin(\phi) + c_i\cos(\phi))^2},\\
 R_{i,\phi}&:=& \frac{\partial R_i}{\partial\phi} = \frac{-d_ic_i\sin(\phi)}{(a_i\cos(\theta)\sin(\phi)+b_i\sin(\theta)\sin(\phi) + c_i\cos(\phi))^2},\nonumber\\
 \alpha_\theta&:= &\frac{\partial\alpha}{\partial\theta}, ~ \alpha_\phi:= \frac{\partial\alpha}{\partial\phi}, ~ \rho'_\theta:=\frac{\partial\rho'}{\partial\theta}=R_{1,\theta} + \alpha_{\theta}(\rho-R_0) - \alpha R_{0,\theta}, \nonumber\\ \rho'_\phi&:=&\frac{\partial\rho'}{\partial\phi}= R_{1,\phi} + \alpha_{\phi}(\rho-R_0) - \alpha R_{0,\phi} ,\nonumber
 \end{eqnarray}


\begin{figure}[h!]
\centering
\includegraphics[scale=0.60]{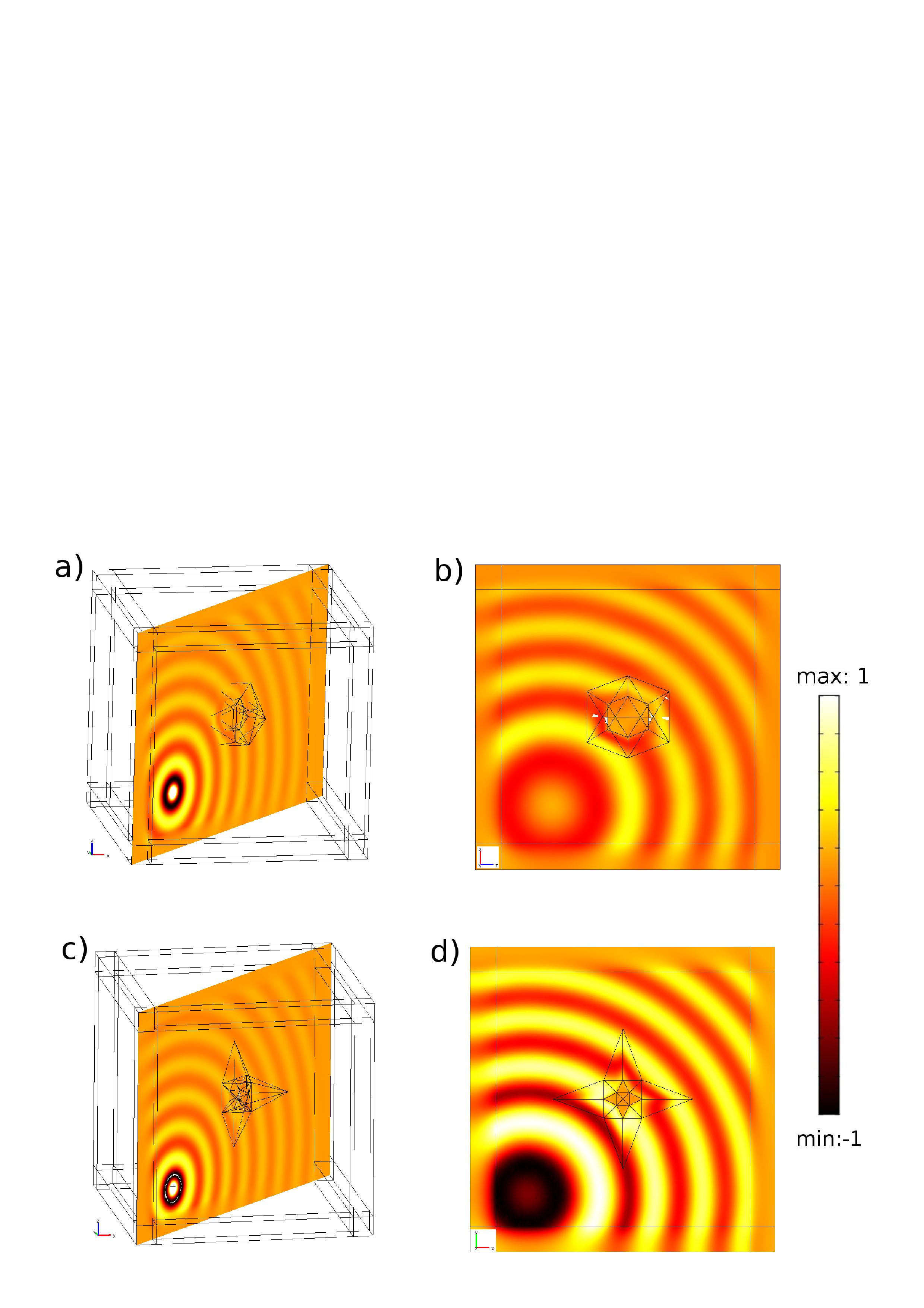}
\caption{\footnotesize ~ Upper panel:
Scattering by an isocahedron cloak in presence of a point source;
(a) Isocahedron cloak 3d-view; (b) Slice in the $(x,z)$-plane for isocahedron with edges $e_1 = 0.2$m and
$e_2 = 0.4$m at the frequencie $\lambda=0.3$m.
Lower panel: (c) Geometry of the 6-point star cloak (3d-view);
(d) Geometry of the (x,y)-plane for a star with edges $e_1 = 0.12$m and  $e_2 = 0.45$m
at the frequency $\lambda = 0.3$m; Lower-Left: Scattering for a point
source in presence of the star cloak (3d-view); Lower-Right:  Scattering for a point
source in presence of the star cloak (2d-view).}
\label{starfig}
\end{figure}

The Jacobian ${\bf J}_{\rho'\rho}$ of the above transformation $(\rho,\theta,\phi)\mapsto (\rho'(\rho,\theta,\phi), \theta'(\rho,\theta,\phi),\phi'(\rho,\theta,\phi)),$ is
\begin{eqnarray}{\bf J}_{\rho'\rho}=\begin{pmatrix} \alpha (\theta,\phi) & \rho'_\theta & \rho'_\phi\\
0&1 &0\\
0&0&1
\end{pmatrix}.\end{eqnarray}

However, for implementation, one may need to get back to Cartesian coordinate system. The following diagram gives the picture as to how to compute the resulting Jacobian ${\bf J}_{x'x}$ of change of coordinates, in Cartesian coordinates.
\[
 \xymatrix{
 (x,y,z) \ar@{->}[rr]^{\displaystyle {\bf J}_{x'x}} \ar@{->}[dd]_{\displaystyle {\bf J}_{\rho x}}
      && (x',y',z') \ar@{->}[dd]^{\displaystyle {\bf J}_{\rho' x'}}    \\ \\
 (\rho,\theta,\phi) \ar@{->}[rr]_{\displaystyle {\bf J}_{\rho'\rho}} && (\rho',\theta',\phi')
 }
 \]

So it now reads
 ${\bf J}_{x'x}={\bf J}_{x'\rho'}{\bf J}_{\rho'\rho}{\bf J}_{\rho x} = {\bf J}_{\phi}diag(1,\rho',\rho'){\bf J}_{\rho'\rho}diag(1,\frac{1}{\rho},\frac{1}{\rho}){\bf J}_{\phi}^{-1}$. We are now ready to deduce the material properties by the usual formula $T^{-1}={\bf J}_{x'x}{\bf J}_{x'x}^{T}/\det({\bf J}_{x'x}).$
 Note that \begin{eqnarray}
{\bf J}_{x'x}{\bf J}_{x'x}^{T}={\bf J}_{\phi}diag(1,\rho',\rho'){\bf J}_{\rho'\rho}diag(1,\frac{1}{\rho},\frac{1}{\rho}) {\bf J}_{\phi}^{-1}{\bf J}_{\phi}^{-T} diag(1,\frac{1}{\rho},\frac{1}{\rho}) {\bf J}_{\rho'\rho}^Tdiag(1,\rho',\rho'){\bf J}_{\phi}^T\end{eqnarray}
and using  ${\bf J}_{\phi}^T{\bf J}_{\phi}= diag(1,\sin^2(\phi),1)$ it becomes
\begin{eqnarray}
{\bf J}_{x'x}{\bf J}_{x'x}^{T}={\bf J}_{\phi}diag(1,\rho',\rho'){\bf J}_{\rho'\rho}diag(1,\frac{1}{\rho},\frac{1}{\rho}) diag(1,\frac{1}{\sin^2(\phi)},1) diag(1,\frac{1}{\rho},\frac{1}{\rho}) {\bf J}_{\rho'\rho}^Tdiag(1,\rho',\rho'){\bf J}_{\phi}^T\end{eqnarray}
so that
${\bf J}_{x'x}{\bf J}_{x'x}^{T}={\bf J}_{\phi}diag(1,\rho',\rho'){\bf J}_{\rho'\rho}
diag(1,\frac{1}{\rho^2\sin^2(\phi)},\frac{1}{\rho^2})
{\bf J}_{\rho'\rho}^Tdiag(1,\rho',\rho'){\bf J}_{\phi}^T.$

This way, the material coefficients will read in Cartesian coordinates
\begin{eqnarray}{\bf T}^{-1}=\frac{1}{det({\bf J}_{x'x})}{\bf J}_{\phi}\begin{pmatrix}
\alpha^2+\frac{\rho'^2_\theta}{\rho^2\sin^2(\phi)}
+\frac{\rho'^2_\phi}{\rho^2}& \frac{\rho'^2_\theta}{\rho^2\sin^2(\phi)} & \frac{\rho'_\phi}{\rho^2}\\
\frac{\rho'_\theta}{\rho^2\sin^2(\phi)} & \frac{1}{\rho^2\sin^2(\phi)}& 0\\
 \frac{\rho'_\phi}{\rho^2} & 0 &  \frac{1}{\rho^2} \end{pmatrix}{\bf J}_{\phi}^T\end{eqnarray}





\subsection{Numerical illustration for a Regular Polyhedron (isocahedron) and its stellated counterpart (6-point star)}

We first consider a linear geometric transformation that maps the point of coordinates $(0,0,0)$ on
an isocahedron $I_1$ of edges of length $e_1=0.2$m and that maps $I_1$ on an isocahedron $I_2$ of
edges of lengths $e_2=0.4$m, ($e_1 < e_2$). One can see in panels (a) and (b) of Figure \ref{starfig}
that the pressure field radiated by an acoustic source is smoothly bent around the inner region
of the cloak (the amplitude of the field indeed vanishes in this region). When the number of edges
increases, the cloak's boundary approximates to certain extent that of a spherical cloak, and the
scattering is reduced, see \cite{diatta2009} for a detailed analysis in the two-dimensional case.
We then consider a linear geometric transformation that maps the points of coordinates $(0,0,0)$ on
a 6-point star $S_1$ of edges of length $e_1=0.12$m and that maps $I_1$ on a 6-point star $S_2$ of
edge of lengths $e_2=0.45$m, ($e_1 < e_2$).  One can see in panels (c) and (d) of Figure \ref{starfig}
that the pressure field radiated by an acoustic source is smoothly bent around the inner region
of the star-shaped cloak. When one increases the number of edges (hence of corners, which become
sharper and sharper), the cloak's boundary becomes
more singular, and the scattering worsens, as shown using an asymptotic approach
in the two-dimensional case in \cite{diatta2009}. Such a behaviour is antagonistic with that of the polyhedra
cloaks.

\section{Concluding remarks}
\label{chap5}
In this paper, we first reviewed the chronology of mathematical results in inverse problems which predate the 2006 proposals of Pendry et al. \cite{pendry} and Leonhardt \cite{leonhardt06} for an invisibility cloak. We trace these results back to 1984, when Kohn and
Vogelius noted the arbitrariness of the Dirichlet-to-Neumann map for conductivity problems involving matrix valued coefficients \cite{kohn84a}, with subsequent works by other mathematicians including Lassas, Uhlmann and Greenleaf who published the first cloaking paper in 2003 \cite{g3}. We reviewed the formulae underlying transformation based acoustic cloaks and performed some finite element numerical simulations for singular and non-singular cloaking effects
in three-dimensional settings. It has been recently shown that acoustic cloaking shells are possible by means
of multi-layered structures consistiing of two types of acoustic isotropic
materials, whose acoustic parameters should be radially varying \cite{sanchez}.
Here, we proposed a design of a broadband multi-layered cloak
with a large number of thin homogeneous isotropic layers, based on a geometric transform for a non-singular
cloak proposed by Kohn et al. \cite{kohn}.
For this, we first rederived the homogenized acoustic 3D equation using
a multiscale asymptotic approach. We found that the homogenized
acoustic parameters are described by a rank-2 tensor (a generalized
density) and a scalar bulk modulus both of which are functions of
the radius. We then performed numerical computations based on the
finite element method which proved that a rigid obstacle surrounded
by a coating consisting of $20$ concentric layers alternating two
types of isotropic bulk moduli and mass densities is neutral
(vanishing backward and forward scattering) for the acoustic plane
waves, over a finite interval of frequencies (in Hertz). The cloak can also mimick the scattering
of a small rigid obstacle.
We finally investigated 3D acoustic cloaks with an irregular star-shape, in order
to exemplify the versatility of transformation acoustics.
We emphasize that our numerical results can have interesting applications in sonar camouflaging in the
context of anamorphism (mirage) effects.

\section*{Acknowledgements}
\label{chap6}
G. Dupont is thankful for a PhD funding from the University of Aix-Marseille III. The authors
wish to thank the referees for many insightful comments.

\end{document}